\def\micron{\hbox{$\mu$m}}
\documentclass{aa}
\usepackage{graphics}

\begin{document}

\thesaurus{3(13.09.1; 13.09.3; 11.17.3; 11.17.1; 11.09.1 PC 1548+437,
 PC 1640+4628, H 0055-259, UM 669, 
B 1422+231, PG 1630+377, PG 1715+535, UM 678; 11.16.1)}
   
\title{ISO continuum observations of quasars at z=1-4
\subtitle{I. Spectral energy distributions of quasars from the UV to far-infrared}
  \thanks{The observations reported here were made with ISO, 
           an ESA project with instruments
           funded by ESA member states (especially the PI countries: France,
           Germany, the Netherlands, and the United Kingdom) and with the
           participation of ISAS and NASA, and with the facilities at the 
           Cerro Tololo Inter-american Observatory (CTIO), NOAO, 
           which is operated by AURA, Inc., under contract with NSF, 
           at the Kiso 
           Observatory, which is operated by Institute of Astronomy, 
           the University of Tokyo, and at the Okayama Astrophysical 
           Observatory, the National Astronomical Observatory of Japan}
}

\author{
  S. Oyabu \inst{1} \and K. Kawara  \inst{1} \and Y. Tsuzuki\inst{1} 
  \and Y. Sofue \inst{1} \and Y. Sato \inst{1} \and H. Okuda \inst{2,3}
  \and Y. Taniguchi\inst{4} \and H. Shibai\inst{5} 
  \and C.~Gabriel \inst{6} \and T. Hasegawa\inst{3} \and E. Nishihara\inst{3}
}

\institute{
  Institute of Astronomy, The University of Tokyo, 2-21-1 Osawa 
  Mitaka, Tokyo, 181-8588, Japan \and 
  Institute of Space and Astronautical Science (ISAS), 3-1-1 Yoshinodai,
  Sagamihara, Kanagawa, 229-8510, Japan \and
  Gunma Astronomical Observatory,  6860-86 Nakayama, Takayama, Agatsuma, Gunma 371-0702, Japan \and
  Astronomical Institute, Tohoku University, Aoba, 
  Sendai 980-8578, Japan \and
  Department of Astrophysics, School of Science, Nagoya University, 
  Furo-cho, Chikusa-ku, Nagoya 464-8602, Japan \and
  ISO Data Centre, Astrophysics Division of ESA, 
  Villafranca, 28080 Madrid, Spain
}

\offprints{S. Oyabu}
\mail{shinki@mtk.ioa.s.u-tokyo.ac.jp}

\date{Received -------- / accepted --------}

\maketitle

\begin{abstract}

Eight luminous quasars with $ -30 < M_B < -27$ at z = 1.4 - 3.7 have been 
observed in the mid- and far-infrared using ISO. All the quasars have been
detected in the mid-infrared bands of ISOCAM, while no far-infrared detections
have been made with ISOPHOT.  
SEDs (Spectral Energy Distributions) from the UV to far-infrared have been 
obtained while supplementing ISO observations with photometry in the optical 
and near-infrared made from the ground within 17 months.
The SEDs are compared with the MED (Mean spectral Energy Distributions) of 
low-redshift quasars with $-27 < M_B < -22$.  
It is shown that our far-infrared observations were limited by confusion 
noise due to crowded sources.

\end{abstract}

\section{Introduction}

The $IRAS$ sky survey was the first observation in which many quasars 
in the far-infrared were detected, and Neugebauer et al.(\cite{neugebauer86}) 
presented $IRAS$ measurements of 179 quasars from 12 to 100 $\mu m$.
Combining $IRAS$ observations with data taken in other wavelengths, 
Sanders et al. (\cite{sanders}) have given SEDs from $\sim$ 0.3 nm
to 6 cm of 109 quasars in the Palomar-Green (PG) survey 
(Green et al. \cite{green}).  Compiling $Einstein$, $IUE$, and
$IRAS$ data with a supplement of ground-based observations, 
Elvis et al. (\cite{elvis}) have presented SEDs for a sample of
47 normal quasars, and derived the SEDs for radio-quiet and radio-loud
quasars. Recently, Andreani et al.(\cite{andreani}) have
presented SEDs for 120 optically selected quasars at low- and high-redshift,
combining sub-mm and mm observations with optical, near-infrared, $IRAS$,
and radio observations. 

The gross shape of quasars SEDs is characterized by two major features; 
the big blue 
bump shortward of 0.3 $\mu m$ and the infrared bump between 2 and 200 $\mu$m 
(Sanders et al. \cite{sanders}; Elvis et al. \cite{elvis}).
It is generally considered that the blue bump is dominated by 
thermal emission from an accretion disk. The infrared bump is ubiquitous in
both radio-quiet and radio-loud quasars.  The infrared emission in 
radio-loud quasars has generally been taken to be dominated by non-thermal
emission (e.g., Impey \& Neugebauer \cite{impey}; Bloom et al. \cite{bloom}; 
Neugebauer \& Matthews \cite{neugebauer99}). 

The origin of the infrared emission in radio quiet quasars is generally
attributed to thermal emission from heated dust. The rise from 1 $\mu$m 
minimum toward 3 $\mu$m, which is universally present in quasars, is naturally 
explained by the sublimation of dust grains at $\sim$ 1,500 K 
(e.g., Kobayashi et al. \cite{kobayashi}).  This is expected from the current 
unified models of active galactic nuclei, which have a dusty obscuring torus 
around a central source.  
Pier \& Krolik (\cite{pier}) computed the infrared properties predicted for 
the dust obscuring tori surrounding central sources by modeling free
parameters of the inner radius of the torus to its thickness, the Thomson 
depths constraining the outer radius of the torus, and the flux of the
nuclear radiation. The infrared emission in 1 - 10 $\mu$m of PG quasars is
well explained by their models, while the emission in the far-infrared is
much greater than that predicted. Several models were proposed to account
for the far-infrared emission of quasars; these are (1)warm dust in a distorted disk 
extending from 0.1kpc to more than 1kpc, which is heated directly by 
radiation from central source (Sanders et al. \cite{sanders}), (2)dust clouds
in the narrow line region heated by the radiation from the central sources 
and star-forming regions (Rowan-Robinson \cite{rowan}), and (3)dust in the
obscuring tori extending to 200 - 3kpc (Andreani et al. \cite{andreani}).
However, it should be noted that temporal variations of 10 $\mu$m
brightness of radio-quiet quasar PG 1535+547 found by Neugebauer \& Matthews 
(\cite{neugebauer99}) casts doubt on the thermal origin of the infrared
emission in radio-quiet quasars. 

High-redshift quasars have been widely recognized to provide unique probes
of high-redshift star formation and galaxy evolution. Metallic abundance
in high-redshift quasars is solar or higher metallicities out to $z > 4$.
Comparing the UV-to-optical spectra of 186 quasars with $0 < z < 3.8$, 
Osmer et al. (\cite{osmer}) concluded that there was no evidence for 
redshift-dependent spectral changes. Hamann \& Ferland (\cite{hamann}) 
analyzed N$_{\mathrm{\ V}}$/C$_{\mathrm{\ IV}}$ and 
N$_{\mathrm{\ V}}$/He$_{\mathrm{\ II}}$ broad emission ratios, and found that 
metallicities in the broad line gas at high redshift are 1 - 10 times solar.
The ratio of UV Fe$_{\mathrm{\ II}}$/Mg$_{\mathrm{\ II}}$ of B 1422+231 at 
$z = 3.6$ shows the the host galaxy was already in the late-evolutionary phase
of the Fe enrichment by SNe Ia at z = 3.6 (Kawara et al. \cite{kawara96};
Yoshii et al. \cite{yoshii}). These all suggest that the central part
of host galaxies formed rapidly at very high-redshift $z \geq 10$, and
metallicity enrichment in the central part was already completed 
at $z = 4 - 5$. 

Dust emission from high-redshift quasars provide another probe to study star 
formation and evolution of host galaxies, especially in
outer regions. Up to date, many quasars at z $>$ 3 have 
been detected in the region from 350 $\mu m$ to 1.3 mm 
(e.g., Andreani et al. \cite{andreani93}; Chini \& Kr\"{u}gel \cite{chini94};
Dunlop et al. \cite{dunlop}; Isaak et al. \cite{isaak};
McMahon et al. \cite{mcmahon}; Ivison \cite{ivison};
Omont et al. \cite{omont96a}; 
Hughes et al. \cite{hughes}; Benford et al. \cite{benford}; 
Carilli et al. \cite{carilli00}). An extensive 240 GHz survey by Chini et al. 
(\cite{chini89}) revealed that the majority of $z < 1$ quasars have
dust masses about a few times 10$^7 M_{\odot}$, comparable to normal spiral
galaxies, thus suggesting that dust is heated by radiation from the central sources.
On the other hand, dust masses $\geq$ 10$^8 M_{\odot}$ have been found in six
of 16 quasars at $z > 4$ (Omont et al. \cite{omont96a}), and CO emission was
found in three of these (Ohta et al. \cite{ohta}; Omont et al. \cite{omont96b};
Guilloteau et al. \cite{guilloteau97},\cite{guilloteau99}; Carilli et al. 
\cite{carilli99}). This may imply that dust emission in
these host galaxies may be dominated by radiation from star-forming regions 
at high-redshift.

To study the distribution of dust in high-redshift quasars, mid- and
far-infrared observations are indispensable. The emission from sublimated dust
could be observed in the mid-infrared, thus probing that the innermost part of the
obscuring tori, and the distribution of dust from obscuring tori to disks (or
outer star-forming regions) could be obtained from far-infrared 
observations.  The $IRAS$ sky survey has only detected  several quasars 
at z $>$ 3 with exceptionally high luminosity in the far-infrared 
(Neugebauer et al. \cite{neugebauer86}; Bechtold et al. \cite{bechtold}; 
Irwin et al. \cite{irwin}); 
F08279+5255 (APM 08279+5255) at z = 3.87, 2126-158 (PKS 2126-15) at z = 3.275, 
and 0320-388 at z = 3.12.  In hoping to detect more quasars, several groups
have carried out mid- and far-infrared photometry using the 
Infrared Space Observatory ($ISO$; Kessler et al. \cite{kessler}).   
The first report of the $ISO$ European
Central Quasar Programme which observed 70 quasars between 4.8 and 200 $\mu m$
including high-redshift quasars has been published by Haas et al. 
(\cite{haas}), and the $ISO/NASA$ AGN Key Project has also observed 72
quasars and AGNs covering a range of redshift up to 4.7
(Wilkes et al. \cite{wilkes}).  
As an $ISO$ open time program, we have executed mid- and 
far-infrared observations of eight quasars at $1.4 < z < 3.7$ using the raster
mapping mode.  This paper presents the results of the photometry of these
quasars supplemented with optical and near-infrared data taken on the ground.
In the course of this work, mid- and far-infrared sources have
serendipitously been discovered. These sources will be described
in the forthcoming paper (Oyabu \& Kawara \cite{oyabu}).
Throughout this paper, $H_0$ = $75km\ s^{-1}\ Mpc^{-1}$ 
with $q_0$ = $0$ is assumed.

\section{Observations and Results}

Eight quasars have been observed with $ISO$ in revolution 169 - 781 
(1996 March - 1998 January).  Optical and 
near-infrared imaging observations have followed within 24 months 
(mostly within 17 months).

\subsection{The Sample}

\begin{center}
\begin{table*}[htbp]
  \caption{Quasars' Sample}
  \begin{tabular}{lrrlllll}
    \hline 
    Object      & \multicolumn{2}{c}{R.A. (J2000) Dec.}       & z & ${M_B}^a$ & Radio$^b$  & Rev (UT)$^c$ &
    Other name and Notes \\
    \hline
    \object{PC 1548+4637} & 15:50:07.6  & +46:28:55  & 3.544    & $-27.0$ & Quiet & 169 (960504) &  \\
    \object{PC 1640+4628} & 16:42:05.1  & +46:22:27  & 3.700    & $-26.8$ & Quiet & 185 (960520) &  \\
    \object{H 0055-2659}  & 00:57:58.1  & $-$26:43:14  & 3.662    & $-29.2$ & Optical & 380 (961130) &  \\
    \object{UM 669}       & 01:05:16.8  & $-$18:46:42  & 3.037    & $-28.4$ & Optical & 415 (970104) 
         & Q 0102-190  \\ 
    \object{B 1422+231}   & 14:24:38.1  & +22:56:01  & 3.62     & $-29.8^d$ & Loud & 424 (970113) & 
      Lensed quasar  \\
    \object{PG 1630+377}  & 16:32:01.1  & +37:37:49  & 1.478    & $-28.2$ & Quiet  & 424 (970113) & 
         Also observed on rev. 778 (980101) \\
    \object{PG 1715+535}  & 17:16:35.4  & +53:28:15  & 1.940    & $-28.5$ & Quiet  & 712 (971027) &  \\ 
    \object{UM 678}       & 02:51:40.4  & $-$22:00:27  & 3.205    & $-29.4$ & Optical & 781 (980104) 
                           & Q 0249-222  \\
    \hline
  \end{tabular}
  \label{tab1}
  \begin{list}{}{}
  \item[$^{a}$] The absolute B magnitude for $H_0 = 75km\ s^{-1}\ Mpc^{-1}$ with $q_0 = 0.0$.
  \item[$^{b}$] Radio property; Quiet = radio quiet, Loud = radio loud, and Optical = optically selected.
  \item[$^{c}$] UT is given in the yymmdd format where yy = year, mm = month, and dd = day.
  \item[$^{d}$] $M_B\sim -26$ after demagnification (Kormann et al. \cite{kormann}).
  \end{list}
\end{table*}
\end{center}

\begin{figure}
     \resizebox{80mm}{!}{\includegraphics{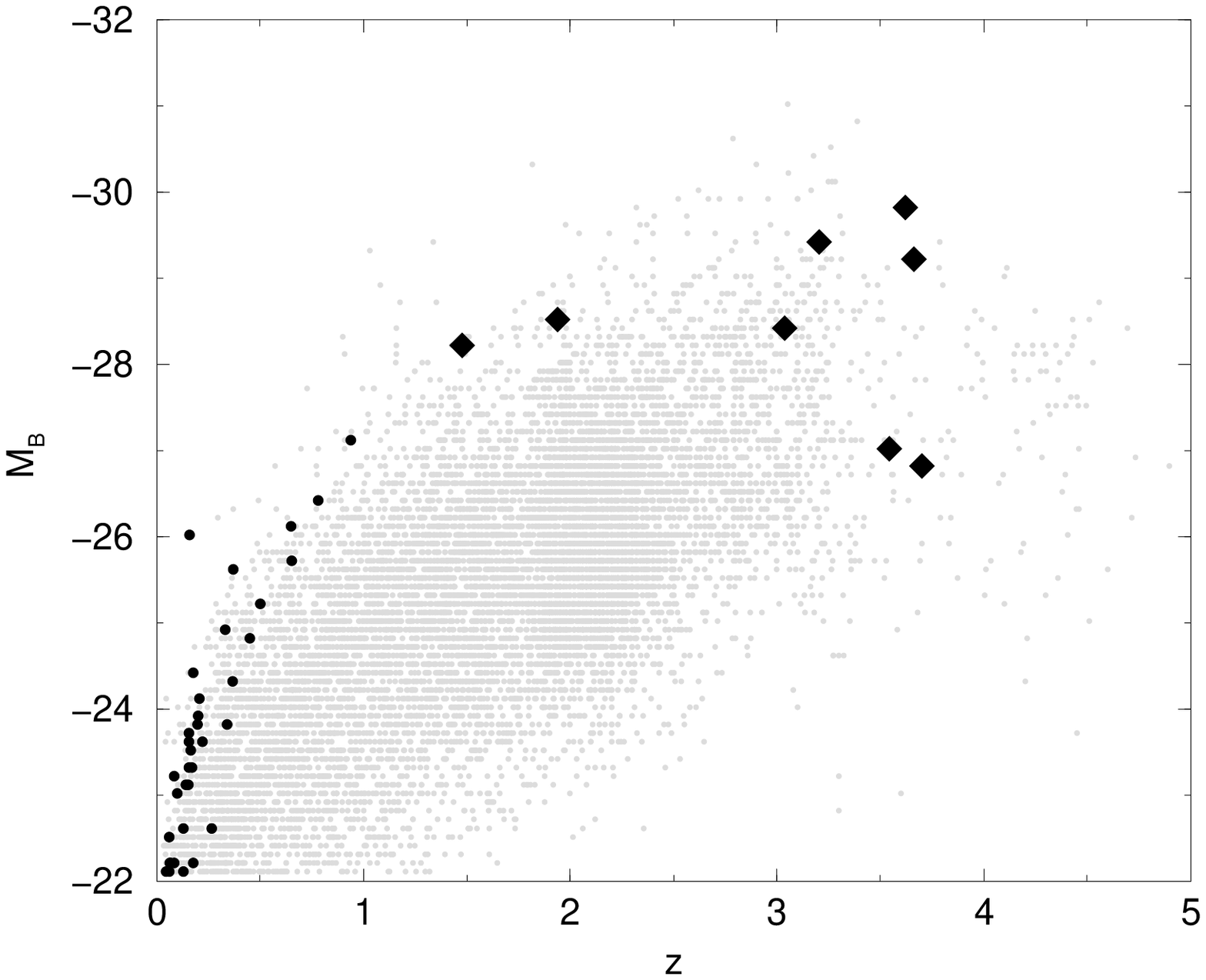}}
     \hfill
     \caption{Sample quasars (large filled diamonds) compared with other 
samples of
quasars on the z - $M_B$ plane. Small filled circles show quasars studied by
Elvis et al.(\cite{elvis}).  Quasars cataloged by 
V\'{e}ron-Cetty \& V\'{e}ron (\cite{veron}) were plotted by dots, which 
appears as a gray background on this plane.}
     \label{fig1}
    \end{figure}

Table \ref{tab1} lists the sample of the eight quasars in the sequence 
of ISO revolutions for execution.  The sample consists of luminous quasars
with $M_B <$ -28 except PC 1548+4637 and  PC 1640+4628; 
these two quasars which were observed at the beginning of this work, turned 
out to be too faint to be detected in the far-infrared, 
and thus the sample 
selection criterion was changed to include very luminous quasars 
in low far-infrared background of infrared cirrus emission.  
Fig. \ref{fig1} presents the eight quasars (large filled diamonds) 
on the $z - M_B$ plane together with those (small filled circles) in the 
sample by Elvis et al.(\cite{elvis}) and those (faint gray points)
complied by V\'{e}ron-Cetty \& V\'{e}ron (\cite{veron}).
All the sample quasars are radio-quiet or optically selected except 
B 1422+231 which is a core-dominant flat-spectrum radio source.\\

\subsection{Mid-infrared observations with ISOCAM}

\begin{table*}
  \caption{Filters and settings for raster mapping}
  \begin{tabular}{lllllll}
    \hline
    \noalign{\smallskip}
Filter$^a$ & LW2  & LW3  &  LW10 &  C\_90 & C\_160   \\  
   \hline 
 $\lambda_{ref}$  & 6.7$\mu$m & 14.3$\mu$m & 12.0$\mu$m & 90$\mu$m & 170$\mu$m \\
 $\Delta\lambda$ & 3.5$\mu$m & 6.0$\mu$m & 7.0$\mu$m & 51$\mu$m & 89$\mu$m \\
 Airy diameter$^b$ & 5.6\arcsec & 12.0\arcsec & 10.1\arcsec &  76\arcsec & 
   143\arcsec   \\    
    \hline \hline
     & \multicolumn{5}{c}{Raster points M $\times$ N,\ \ Raster step,
     \ \ Exposure time per raster point$^c$,\ \ Pixel field of view} \\
    \cline{2-6}
QSO field     & LW2     & LW3   &  LW10  &  C\_90 & C\_160  \\

    \hline
    \noalign{\smallskip}
   PC 1548+4637 
    & $5 \times 2$,\  6\arcsec,\ 60s,\    6\arcsec & ... 
    & $4 \times 1$,\  6\arcsec,\ 25s,\    6\arcsec 
    & $4 \times 4$,\ 44\arcsec,\ 64s,\ 44\arcsec 
    & \multicolumn{1}{r}{$3 \times 3$,\ 90\arcsec,\ 64s,\ 89\arcsec}  \\
   PC 1640+4628
    & $5 \times 2$,\  6\arcsec,\  60s,\    6\arcsec & ... 
    & $4 \times 1$,\  6\arcsec,\  25s,\    6\arcsec 
    & $4 \times 4$,\ 44\arcsec,\  64s,\ 44\arcsec 
    & \multicolumn{1}{r}{$3 \times 3$,\ 90\arcsec,\  64s,\ 89\arcsec} \\
   H 0055-2659
    & $5 \times 2$,\  6\arcsec,\  60s,\    6\arcsec & ...  & ...
    & $5 \times 3$,\ 44\arcsec,\  69s,\ 44\arcsec 
    & \multicolumn{1}{r}{$4 \times 2$,\ 90\arcsec,\  73s,\ 89\arcsec} \\
   UM 669
    & $  6 \times 6$,\  7\arcsec,\  90s,\    3\arcsec & ...  & ... & ... 
    & \multicolumn{1}{r}{$10 \times 10$,\ 46\arcsec,\  16s,\ 89\arcsec} \\
   B 1422+231
    & $  4 \times 4$,\  7\arcsec,\  90s,\    3\arcsec 
    & $  4 \times 4$,\  7\arcsec,\  90s,\    3\arcsec  & ... & ... 
    & \multicolumn{1}{r}{$10 \times 10$,\ 46\arcsec,\  16s,\ 89\arcsec} \\
   PG 1630+377$^d$
    & $  4 \times 4$,\  7\arcsec,\  38s,\    3\arcsec 
    & $  4 \times 4$,\  7\arcsec,\  38s,\    3\arcsec  & ... & ... 
    & \multicolumn{1}{r}{$10 \times 10$,\ 46\arcsec,\  16s,\ 89\arcsec} \\
   PG 1715+535
    & $  4 \times 4$,\  7\arcsec,\  38s,\    3\arcsec 
    & $  4 \times 4$,\  7\arcsec,\  38s,\    3\arcsec  & ... & ... 
    & \multicolumn{1}{r}{$10 \times 10$,\ 46\arcsec,\  16s,\ 89\arcsec} \\
   UM 678
    & $  4 \times 4$,\  7\arcsec,\  90s,\    3\arcsec 
    & $  4 \times 4$,\  7\arcsec,\  90s,\    3\arcsec  & ... & ... 
    & \multicolumn{1}{r}{$6 \times 5$,\ 92\arcsec,\  64s,\ 89\arcsec} \\
    \noalign{\smallskip}
    \hline          
  \end{tabular}
  \label{tab2}
  \begin{list}{}{}
  \item[$^{a}$] Cited from $ISO\ Handbook\ Volume\ III\ (CAM)$ 
  (Siebenmorgen et al. \cite{siebenmorgen}) 
  and $Volume\ V\ (PHT)$.
  \item[$^{b}$] The aperture photometry was performed by using the apertures
       with a diameter of $2\times d_{airy}$ for ISOCAM  and of
       $d_{airy}$ for ISOPHOT. The diameter of the Airy disk $d_{airy}$
       is $2.44(\lambda/60cm)$ in radian or 
      $0.84\lambda(\mu m)$ in arcsec.  Note that the FWHM of 
                the Airy disk is $0.422 \times d_{airy}$.
  \item[$^c$]  ISOCAM exposure time per raster point is given as 
$T_{int}\times N_{exp}$ where $T_{int}$ is an integration time of a single 
exposure and $N_{exp}$ is the number of exposures. $T_{int}= 5 sec$ was used 
for all LW2 and LW3 observations except for PG 1630+377 and PG 1715+535 for
which $T_{int}=2 sec$ was used. $T_{int}=2 sec$ was used for LW10 observations.
  \item[$^{d}$] The second observation was executed on revolution 778 in
                C\_160 with a parameter set of 
                ($6 \times 6$,\ 92\arcsec,\ 64s, 89\arcsec).
  \end{list}
\end{table*}
\begin{table*}
  \caption{ISO mid/far-infrared flux density and UV to IR luminosity}
  \begin{tabular}{lrrrcrrcrrr}
    \hline
    & \multicolumn{3}{c}{ISOCAM(mJy)} & & \multicolumn{2}{c}{ISOPHOT(mJy)}
     & & \multicolumn{3}{c}{Luminosity$(10^{13}L_{\sun})^a$}\\ 
    \cline{2-4} \cline{6-7} \cline{9-11}
  Object      & LW2               & LW3 & LW10  & & C\_90 & C\_160
    & & $L_{UVO}$ & $L_{ir}$ & $L_{1.25\mu m}$ \\ 
    \hline
  PC 1548+4637 & $0.16\pm0.04$ & ...   & $0.67\pm0.17$ & & $8.2\pm28$ 
               & $32\pm80$ & & 11 & $<130$ & 1.7\\
  PC 1640+4628 & $0.15\pm0.04$ & ...   & $0.29\pm0.11$ & & $-8.3\pm20$ 
               & $83\pm121$ & & 8.1 & $<240$ & 1.8  \\
  H 0055-2659  & $0.29\pm0.06$   & ...           &  ...        & & $16\pm19$ 
               & $-2.7\pm60$  & & 24 & $<160$ & 3.3 \\
  UM 669       & $0.50\pm0.01$   & ...           &  ...        & &  ...       
               & $-9.9\pm18$ & & 27 & $<39$ & 3.5 \\
  B 1422+231   & $5.8\pm0.06  $   & $15.1\pm0.07 $ & ...         & &  ...     
               & $83\pm56$ & & 410 & $<320$ & 64   \\
  PG 1630+377  & $4.3\pm0.04  $   &  $7.3\pm0.07$   & ...         & & ...      
               & $5.6\pm14$$^b$ & & 28  & $<16$ & 4.3 \\
  PG 1715+535  & $4.7\pm0.06  $   &    $10.3\pm0.06$   & ...      & & ...      
               & $-13\pm16$  & & 81  & $<26$ & 9.4 \\
  UM 678       & $0.73\pm0.01$   &   $1.9\pm0.03$   &  ...        & &  ...     
               & $-15\pm58$ & & 30   & $<116$  & 5.9 \\
    \hline
  \end{tabular}
  \label{tab3}
\begin{list}{}{}
\item[$^a$] The UVO ($0.1\mu m$ to $1 \mu m$) luminosity $L_{UVO}$ = 
 $4\pi D_L^{2} \times 2.3\int_{14.5}^{15.5}\nu f_\nu d\log_{10}\nu$,  
 the IR ($1.5\mu m$ to $100 \mu m$) luminosity $L_{ir}$ = 
 $4\pi D_L^{2} \times 2.3 \int_{12.5}^{14.3}\nu f_\nu d\log_{10}\nu$, 
 and the 1.25 $ \mu m$ luminosity $L_{1.25\mu m}$ = $4\pi D_L^{2}\nu f_\nu$,
 where $D_L$ and $f_{\nu}$ denote the luminosity distance and observed
 flux density, respectively. Where no observations exit such 
 as $1 \mu m$ flux denisities, interpolated flux densities were used. 
 $L_{UVO}$ is based on the optical and near-infrared flux densities given in 
 Table \ref{tab4}. 
\item[$^b$] $f_\nu(174\mu m) = 89\pm78$ mJy was obtained 
   in the second observation.
 \end{list}

\end{table*}
\begin{figure}
  \resizebox{80mm}{!}{\includegraphics{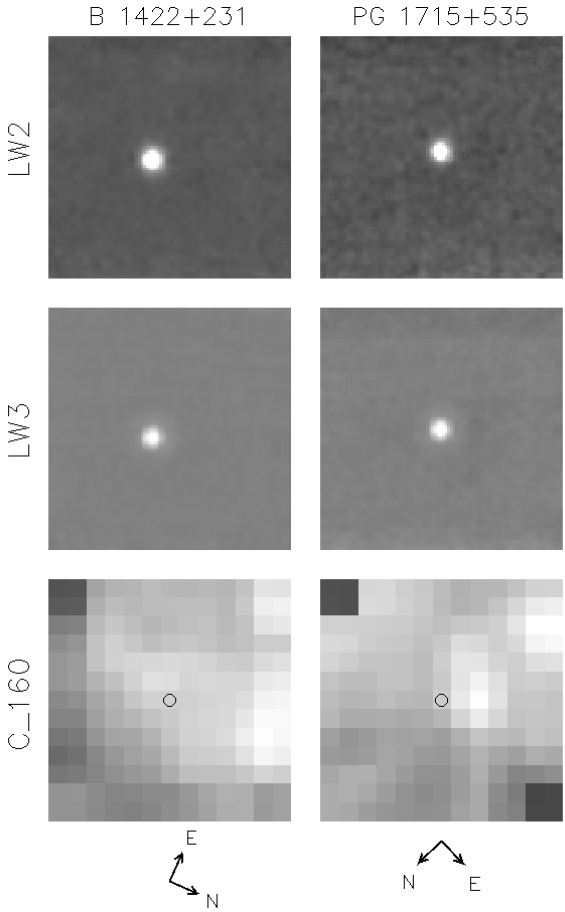}}
  \hfill
  \caption{ISOCAM LW2(6.7\micron) and LW3(14.3\micron) and ISOPHOT 
    C\_160(170\micron) maps are shown on the top, middle, and bottom 
    panels, respectively. B 1422+231 shown on the left was observed on 
    revolution 424 and PG 1715+535 on revolution on 712. The ISOCAM and ISOPHOT
    maps are 90\arcsec and 598\arcsec(10\arcmin), respectively, with North 
    and East as indicated by the arrows. Quasars should be at the 
    center of the gray circles.}
  \label{map}
\end{figure}
The mid-infrared observations were performed with ISOCAM 
(Cesarsky et al. \cite{cesarsky}).  Three broad band filters, namely, 
LW2 (reference wavelength 6.7\micron), LW3(14.3\micron), and LW10 
(12.0\micron) were used. All the quasars were observed in LW2 with additional
measurements in LW3 or LW10.  To detect faint sources against the intense 
background dominated by zodiacal light, the AOT (Astronomical Observation 
Template) CAM01 which is the microscan raster mapping mode was used to achieve 
accurate flat-fielding (e.g., Taniguchi et al.~\cite{TA97}; Altieri et 
al.~\cite{AL99}).  Table \ref{tab2} lists the details of CAM01 parameters as well 
as the characteristics  of the broad band filters.  

The standard ISOCAM reduction software CIA 3.0\footnote{CIA is distributed 
by the ISO Data Centre and the ISOCAM Data Centre at Saclay, France.} was 
used to produce ISOCAM images from ERD (Edited Raw Data). This process
includes dark subtraction, deglitching, correction for the transient
behavior of ISOCAM pixel signals, and flat fielding(Delaney \cite{delaney}).
The inversion transient correction model of Starck et al.(\cite{starck}) was 
applied. The factor of the correction for the transient
behavior is 0.58 - 0.77 in LW2, 0.79 - 1.0 in LW10,
and 0.87 in LW10.
Fig.~\ref{map} shows LW2 and LW3 maps for the two brightest quasars B 1422+231 
and PG 1715+535. All the quasars were clearly detected at the expected 
positions. 
Aperture photometry was performed using IDL. 
Two apertures centered on the object were used; the small one has a diameter 
of $2d_{airy}$, two times the Airy diameter as given in Table \ref{tab2} and 
the other has a diameter of $4d_{airy}$. The photometry was corrected for 
loss of flux in the
PSF (Point Spread Function) wings by computing the PSF based on the model 
having a two mirror f/15 telescope with radii for the primary
and secondary mirror of 30 and 10 cm, respectively(M\"{u}ller ~\cite{muller}).
The factor of the PSF correction is 0.75 - 0.90 (i.e., loss of flux is 
0.1 - 0.25), depending on the raster step 
and the pixel field of view. 
 
The results after these corrections are given in Table \ref{tab3} with 
statistical errors. The errors in the absolute photometric calibration are not included in Table 3; these errors are estimated to be 15\%(Siebenmorgen et al. \cite{siebenmorgen}).

\subsection{Far-infrared observations with ISOPHOT}

The far-infrared observations were made with ISOPHOT 
(Lemke et al. \cite{lemke}). All the quasars were observed in the broad
C\_160 (170\micron) band with additional measurement in C\_90 (90\micron).
ISO far-infrared surveys (Kawara et al. \cite{kawara98}; Puget et al. 
\cite{puget}) clearly indicate that the sky seen in the far-infrared has a 
clumpy structure which is made up of IR cirrus and 
extragalactic sources.  This structure rotates with time relative to the ISO 
coordinate system due to the field rotation, increasing the probability of
fault detection if the chopping mode is used. 
We thus selected the PHT22 staring raster map mode to make small maps around
the quasar. Table \ref{tab2} presents the characteristics of the 
photometric filters and the details of observational parameters for 
raster mapping.

ISOPHOT images were produced by using the standard ISOPHOT reduction software 
PIA V7.3 and V8.1\footnote{PIA is a joint development by the ESA Astrophysics 
Division and the ISOPHOT consortium.} (Gabriel et al. \cite{gabriel}), starting
at the edited raw data (ERD) created via the off-line processing 
version 7.0. The AOT/Batch processing mode of PIA is used with the default
parameters to reduce ERD to the Astronomical Analysis Processing 
(AAP) level. This standard reduction includes linearization and deglitching of 
integration ramps on the ERD level, signal deglitching and drift recognition
on the SRD (Signal per Ramp Data) level, reset interval normalization, signal 
deglitching, dark current subtraction, signal linearization, and
vignetting correction\footnote{Vignetting correction is set to on in the default setting of PIA. However, the effect of the correction is none; just multiplying the data by 1.0, because the chopper throw is zero in the PHT22 raster mode.} on the SCP(Signal per Chopper Plateau data)
level. The responsivity calibration was made on the SPD (Standard Processed
Data) level by using the second measurement of the internal Fine Calibration 
Source 1(FCS1) which is calibrated against celestial standards.  The correction
for the transient behavior of the detectors was applied to 
point sources (quasars) on this level.  Images were produced on the AAP 
(Astronomical Analysis Processing) in the mapping mode with median 
brightness values. 

The correction for drift in the responsivity is not important to our
observations, and so this correction was not made. 
To check the importance of the drift, the MEDIAN filter 
technique was applied to the results from AAP (hereafter called AAP map).
Applying this technique to large AAP maps in the Lockman hole, Kawara 
et al. (\cite{kawara98}) show that this is a powerful tool to correct for 
drift in the detector responsivity.  However, unlike large AAP maps in the
Lockman hole, the MEDIAN filter technique does not improve our AAP maps.
This is attributed to the size of our AAP maps; the observing time of
these small maps is shorter than the timescale of drift in the detector 
responsibility, and so impact by the drift is small. In fact, every detector 
pixel of the 
C100 and C200 detector arrays has been checked for signal, and neither spike 
noise nor drift in the responsibility was found. Maps with all the detector
pixels always give better results than those obtained by masking some of
the detector pixels. In addition, it was confirmed that there were no 
significant differences between two AAP maps produced by two different
algorithms on the AAP level, namely, the full coverage and distance 
weighting algorithms. Fig.~\ref{map} shows AAP maps of C\_160 for the 
brightest quasars B 1422+231 and PG 1715+535.

Aperture photometry was then performed using IRAF\footnote{IRAF is 
distributed by NOAO.} and Skyview\footnote{Skyview is distributed by IPAC.} 
in the manner 
similar to ISOCAM.  The original pixel sizes of AAP maps are equal to
the raster steps.   Because these are too large to center the aperture on the 
quasar with sufficient accuracy, AAP maps were rebinned in such a way that 
each original pixel is converted into  $10 \times 10$ sub-pixels. 
Two apertures centered on the object were used; the small one has a diameter 
of $d_{airy}$, the Airy diameter as given in Table \ref{tab2}, and 
the other has a diameter of $2d_{airy}$. The photometry was corrected for loss of flux in the
PSF (Point Spread Function) wings by computing the same PSF model as used for
ISOCAM. The factor of the PSF correction is 0.63 except for PC 1548+4632 and
PC 1640+4628. These two quasars were centered between four pixels in such a 
way that quasars illuminate the four pixels equally. 
Consequently the loss of flux measured with the two 
apertures is large, and the factor of the  PSF correction is 0.27 for C\_90
and 0.23 for C\_160. 

The results after these corrections are given in Table \ref{tab3} with 
statistical errors. The errors in the absolute photometric calibration 
are not included in Table \ref{tab3}; these errors are estimated to be 30\% 
(Klaas et al. \cite{klaas00}).
It is noted that PG 1630+377 was observed twice at $170\mu m$ to check 
variability.  

\subsection{Optical and near-infrared data from ground-based observations}

Optical images were taken on the 0.9m telescopes at CTIO and the Schmidt 
1.05m telescope at Kiso Observatory.  Near-infrared imaging was made 
in the standard dithering mode on the 1.88m telescope at the Okayama 
Astrophysical Observatory, NAOJ. SExtractor (Bertin \& Arnouts 
\cite{sextractor}) was applied to the optical and near-infrared images 
to perform photometry. 
Flux calibration was made using the standards given by Landolt (\cite{landolt})
in the optical and the UKIRT faint standards (Casali et al. \cite{ukirt}) in
the near-infrared.  A typical photometric error is 0.05 mag. 

Table \ref{tab4} presents magnitudes of the quasars together
with statistical errors.  As shown in
the table, all the ground-based observations were performed within 24 months 
(mostly 17 months) 
from the ISO observations so as to reduce the chance of having flux 
variations 
between ISO and ground-based observations. Table \ref{tab4} also 
supplements optical magnitudes taken by 
Richards et al. (\cite{richards}) and Schneider et al. (\cite{schneider}). 
All the supplementary quasars but PG 1548+4637 were observed within 27 months 
before the ISO observations.

\begin{table*}
  \caption{Optical to near-infrared magnitudes}
  \begin{tabular}{lllllllll}
    \hline
    \multicolumn{9}{c}{This work} \\
    \hline
    Object      & V              & R              & I              & J  & H  
     &  K$^\prime$$^a$  & UT$^b$ & Obs.$^c$ \\
    \hline
    PC 1640+4628 & $20.13\pm0.12$ & $19.38\pm0.16$ & $18.80\pm0.15$ 
       & ... & ... & ... & 970828 & Kiso \\
    PC 1640+4628 & ...   & ...            & ... 
       & ... & $17.8\pm 0.2$ & ... & 980606 & OAO  \\
    H 0055-2659  & $18.17\pm0.02$ & $17.79\pm0.02$ & $17.69\pm0.03$ 
       & ... & ... & ... & 971018 & CTIO \\
    UM 669  & $17.65\pm0.01$ & $17.47\pm0.01$ & $17.23\pm0.02$ 
       & ... & ... & ... & 971018 & CTIO \\
    PG 1630+377  & $16.13\pm0.01$ & $15.76\pm0.16$ & $15.51\pm0.15$ 
       & ... & ... & ... & 970824 & Kiso  \\
    PG 1630+377  & ... & ... & ...  
       & $14.92\pm0.03$ & $14.38\pm0.01$ & $14.12\pm0.04$ & 980603 & OAO \\
    UM 678  & $17.63\pm0.02$ & $17.62\pm0.02$ & $17.30\pm0.03$ 
      & ... & ... & ... & 971018 & CTIO \\
    \hline \hline 
    \multicolumn{9}{c}{Data from others} \\
   \hline
  Object$^d$ & u$^\prime$ & g$^\prime$ & r$^\prime$  & i$^\prime$ & z$^\prime$
               & \multicolumn{2}{r}{UT} & Obs.$^e$ \\
  \hline
  PG 1715+535  & 16.53 & 15.89 & 15.47 & 15.25 & 15.24 
               & \multicolumn{2}{r}{July 1995} & R \\
  PG 1630+377  & 16.27 & 16.13 & 15.94 & 15.83 & 15.70 
               & \multicolumn{2}{r}{July 1995} & R \\
  B 1422+231   & ...   & 16.33 & 15.18 & 15.07 & ...        
               &  \multicolumn{2}{r}{July 1995} & R \\      
  PC 1548+4637 & ...   & ... & 19.27$^f$ & ...  & ... 
               &  \multicolumn{2}{r}{April 1987} & S  \\
  \hline
  \end{tabular}
  \label{tab4}
  \begin{itemize}
  \item[$^a$] The K$^\prime$ bandpass filter ($2.15\pm0.15\mu m$) is
      slightly narrower and bluer than 
      the standard K filter ($2.20\pm0.20 \mu m$).
  \item[$^b$] Universal time when observed in the yymmdd format.
  \item[$^c$] At the Kiso Observatory, the 2048$^2$ CCD with
              1.5\arcsec per pixel was used on the 105cm Schmidt telescope. \\
              At CTIO, the 2048$^2$ CCD with 0.4\arcsec per pixel was used 
              on the 90cm telescope. \\
              At OAO (Okayama Astrophysical Observatory), the infrared imager
              spectrometer which has a HgCdTe 256$^2$ detector array with 
              0.97\arcsec per pixel was used on the 188cm telescope.             \item[$^d$] All but PG 1548+4637 were observed within 27 months before the
              ISO observations.
  \item[$^e$] R = Richards et al. (\cite{richards}); 
              S = Schneider et al. (\cite{schneider}).
  \item[$^f$] The $r_4$ bandpass was used.
  \end{itemize}
\end{table*}

\section{Discussion}

\begin{figure*}[htbp]
  \resizebox{18cm}{!}{\includegraphics{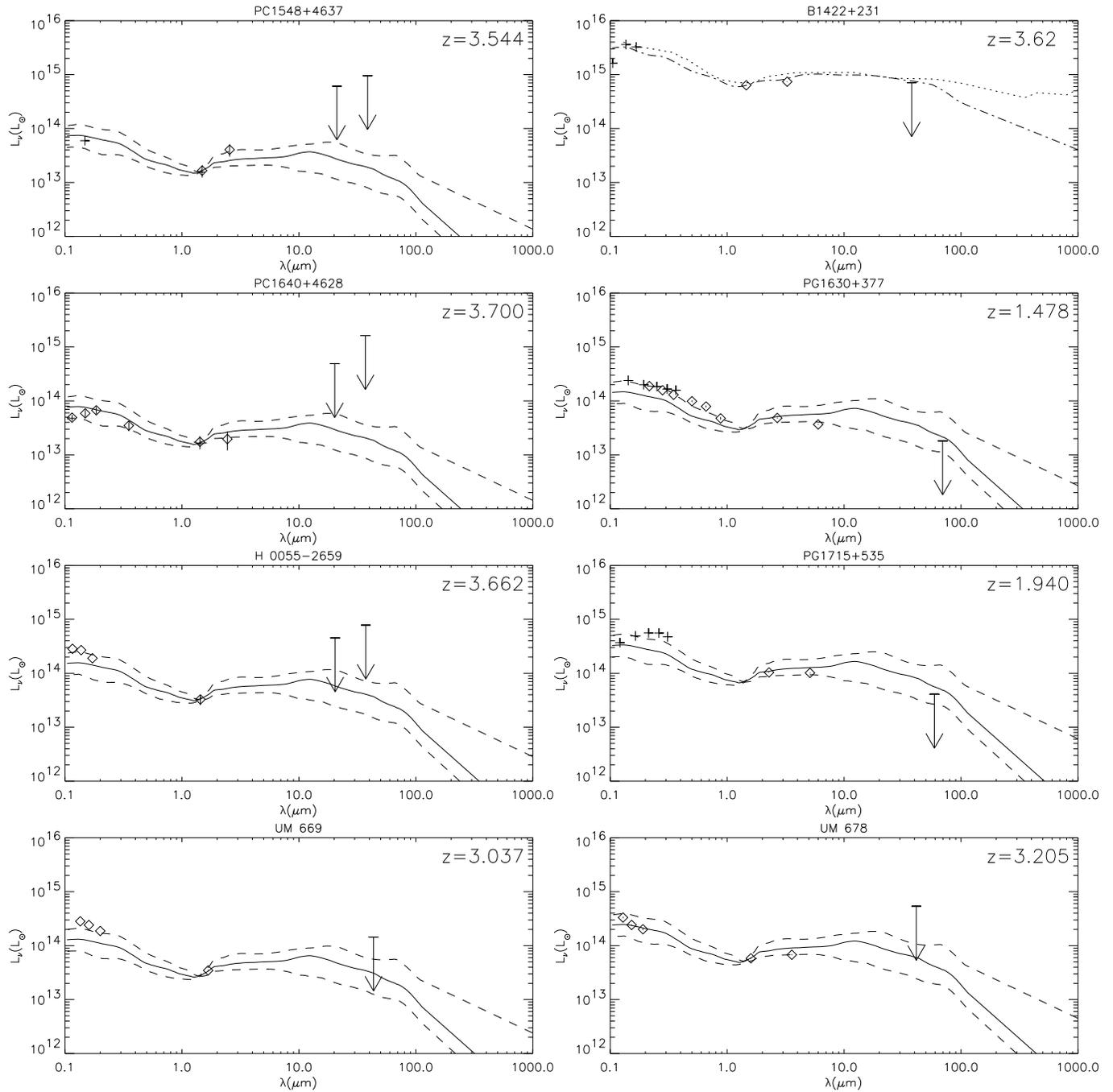}}
  \hfill
  \parbox[b]{18cm}{
    \caption{Restframe SEDs compared with the MED of low-redshift
quasars compiled by Elvis et al. (\cite{elvis}). Dashed lines show the 1 
$\sigma$ 
deviation (actually 68 Kaplan-Meier percentile) envelopes.
Asterisks denote data taken by Richards et al.(\cite {richards}) and 
Schneider et al.(\cite {schneider}). The dot-dashed and dot lines for 
B 1422+231 shows
the MED of radio-loud quasars from Elvis et al.(\cite{elvis}) and 
the SED of 3C273 from Lichti et al.(\cite{lichti}), respectively. 
Note that 3 $\sigma$ upper limits correspond three times uncertainties 
listed in Table \ref{tab3}.
}
    \label{fig:sed}}
\end{figure*}

All the quasars have been detected in the ISOCAM LW2 and LW3 or LW10 band.
Combining optical and near-infrared observations, SEDs from the UV to
mid-infrared have been obtained for eight quasars at z = 1.4 - 3.7. 
Fig. \ref{fig:sed} shows the restframe SEDs ($L_{\nu}$\footnote{$L_{\nu}$ 
is defined by $4\pi D_L^{2} \nu f_{\nu}$ where $D_L$ denotes the luminosity 
distance to the object.}). The down arrows indicate the 3 $\sigma$ upper 
limits obtained with the ISOPHOT observations. Note that 3 $\sigma$ upper 
limits in Fig. \ref{fig:sed}  correspond to three times uncertainties 
listed in Table \ref{tab3}.
The mean spectral distributions
(MEDs) of low redshift quasars by Elvis et al.(\cite{elvis}) are also shown
by fitting to the $ISO$ $6.7\mu$m flux density. Note that
$6.7\mu$m in the observing frame is approximately $1.25\mu$m in the restframe.
The dash lines indicate the 1 $\sigma$ deviation (strictly speaking 68 
Kaplan-Meier percentile) envelopes. The rise from the 1 $\mu$m minimum toward
3 $\mu$m is evident at least in $z = 3.5$ quasar PC 1548+4637. This rise is 
naturally explained by the sublimation of dust grains at $\sim$ 1,500 K 
(e.g., Kobayashi et al. \cite{kobayashi}) as expected from  the current 
unified models of active galactic nuclei.  Hence, our data suggest that the 
obscuring torus model could be applied to high-redshift quasars.

The original objective of this work was to study whether there was any evidence
for redshift-dependent changes in SEDs from the UV to far-infrared. As shown
in Fig. \ref{fig:sed}, the data from the UV to mid-infrared do not
deviate significantly from the MED of the low-redshift sample.
However, this cannot be taken as evidence for no SED evolution in the UV to 
mid-infrared, because our sample is too small in number and too narrow in 
spectral coverage.
Our far-infrared observations only provide 3 $\sigma$ upper limits, and are
not very helpful to check SED evolution in the far-infrared. 

The far-infrared detection with $ISO$ have been severely limited by the 
confusion 
due to galaxies and local peaks of the IR cirrus (Kawara et al. 
\cite{kawara98}; Herbstmeier et al. \cite{herbstmeier}; 
Matsuhara et al. \cite{matsuhara}).  Simulating the 90 and 170 $\mu$m source 
counts  in the Lockman Hole, Kawara et al. (\cite{kawara00a}, \cite{kawara00b})
concluded that the effect of confusion due to crowded sources (presumably 
extragalactic) was significant in the flux 
range below 200 mJy, and 50\% of 140 mJy sources at 170 $\mu$m were 
left undetected 
due to confusion noise.  The expected fluxes of our sample quasars ranges 
from 7 - 170 mJy at 170 $\mu$m.  Such confusion is clearly seen in  
Fig.~\ref{map}.  There are a few sources scattering around the quasar, 
which hampers detection in the far-infrared. It should be noted that the 
brightest source in the PG 1630+377 field has 106 mJy and the one in 
PG 1715+535 has 190 mJy (Oyabu \& Kawara \cite{oyabu}).  

It is clear that larger aperture telescopes like FIRST, which provide
spatial resolution finer than ISO, are required for far-infrared studies of 
high-redshift quasars. In cases where similar aperture telescopes such as
IRIS(ASTRO-F) and SIRTF are used, data sampling, namely a step size of raster 
mapping, 
should be carefully designed so that far-infrared images can be improved
by deconvolution without violating the sampling theorem
(Magain, Courbin, \& Sohy \cite{magain}). 

\acknowledgement We wish to thank the staff of Vilspa, CTIO, OAO, and the Kiso
observatory for their assistance and hospitality.  We are grateful to T.Tsuji
for his continuous support to this work. We thank the referee, P. Andreani, 
for valuable suggestions on how to improve this work.

\end{document}